\documentclass{article}

\usepackage{algorithmic}
\usepackage{algorithm}
\usepackage{graphicx}
\input{epsf.sty}

\renewcommand{\equiv}{\sim}

\newcommand{\set}[1]{\{ #1 \}}

\newcommand{\inv}{\preceq}

\newcommand{\F}{{\cal F}}

\newtheorem{definition}{Definition}
\newtheorem{lemma}[definition]{Lemma}
\newtheorem{theorem}[definition]{Theorem}

\newtheorem{proposition}[definition]{Proposition}
\newtheorem{corollary}[definition]{Corollary}

\newtheorem{question}{Question}

\newcommand{\qed}{\hfill \rule{1ex}{1ex}} %right justified box

\newenvironment{proof}{{\bf Proof}: }{\qed}

\title{Sorting with a Forklift}
\author{
M.~H.~Albert
\and
M.~D.~Atkinson
}

\begin{document}

\maketitle

\begin{abstract} 
A fork stack is a generalised stack which allows pushes and pops of
several items at a time. We consider the problem of determining which
input streams can be sorted using a single forkstack, or dually, which
permutations of a fixed input stream can be produced using a single
forkstack.  An algorithm is given to solve the sorting problem and the
minimal unsortable sequences are found.  The results are extended to
fork stacks where there are bounds on how many items can be pushed and
popped at one time. In this context we also establish how to enumerate
the collection of sortable sequences.
\end{abstract}

\section{Introduction}
\label{INTROSEC}
There is a close historical connection between the investigation of
permutation classes closed under pattern containment, and the study of
what sequences can be generated (or sorted) using a particular data
structure (see for example \cite{Atk2}, \cite{Knu}, \cite{Pra},
\cite{Tar}). Indeed it could be argued that the parents of the study
of these permutation classes are the Erd\"{o}s-Szekeres theorem and
Knuth's result that the permutations sortable with a single stack are
precisely those which do not contain 231 as a pattern.

One of the purposes of a data structure is to accept input data, store
it in some form, and then release it in response to certain
requests. In most structures these functions may be interleaved with
one another. If we observe only the order that data is input to a data
structure, and then the order in which it is released, the operation
of the data structure will simply be perceived as generating a
permutation of the data. So it is natural to associate with a data
structure the collection of such permutations which it can
realize. Furthermore, most natural data structures have a hereditary
property. That is, if they can achieve a certain permutation of a
large collection of data, then they can achieve the restriction of
that permutation to any subset of the data. Subject to this property,
the permutations associated to a data structure will be a class of
permutations closed under pattern containment.

The investigation of classes connected to data structures in this way
is greatly facilitated by keeping this connection in mind. Generally
speaking, thinking about how the data are assigned to storage,
manipulated within storage, and released from storage, will allow a
clearer understanding of the corresponding class of permutations. In
this paper, we carry out this program with respect to a new data
structure, the {\em forkstack}, a stack in which it is possible to add
or remove multiple data items with a single operation.

The stack is an ubiquitous data structure, used in many algorithms,
typically where last in, first out, behaviour is required or
desirable. In some contexts however, the standard stack structure with
its limited push and pop operations can seem overly
restrictive. Consider, for example, the situation where there are two
stacks, each containing a sorted sequence of data values with smallest
elements on top. It is desired that these sequences be merged into a
single sorted sequence. If this merge is to take place into a new
array, or into a queue, then the traditional stack is perfectly suited
for the task. However, for reasons of parsimony or elegance, one might
wish to accomplish this merge in place. In a standard stack, this is
difficult. However, if we could pop (and push) sequences of elements
from the top of each stack, then it becomes simplicity itself. One
simply pops from the stack with smaller head the maximal sequence
which ends with an element smaller than the head of the other stack,
and pushes this sequence onto the other stack. This process is
repeated until one of the stacks is empty. Note that if the stacks are
implemented as linked lists, then this entire process is simply a
matter of repeated pointer assignments. Also, by keeping track of the
location of the original larger head the whole process can be
accomplished by a single pass through the stacks.

Since the operations of a forkstack are more flexible than those of an
ordinary stack it will be prudent to provide an algorithm for carrying
out the process of sorting an input permutation, as well as an
abstract characterisation of obstructions to sortability. We will
also, in some cases, be able to explicitly determine algebraic
relations satisfied by the generating functions that enumerate the
number of sortable sequences of each length using push and pop
operations which are bounded in size. We will describe a method which,
in principle, allows all such enumeration problems to be resolved.

\begin{figure}[ht]
\begin{center}
\epsfysize=4cm
\epsffile{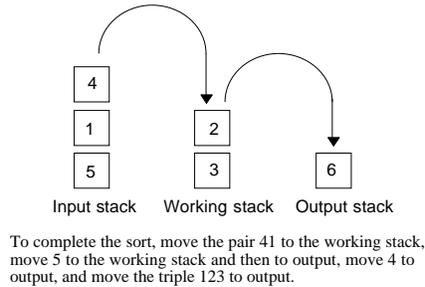}
\caption{A snapshot of sorting}
\label{fork}
\end{center}
\end{figure}

The name, {\em forkstack}, that we have given this data structure is
derived from the following analogy which we find sufficiently
powerful that it forms the foundations of our understanding of the
structure. Begin with a stack of boxes, called the input, labelled
1 through $n$ in some order. A powerful
forklift can remove any segment of boxes from the top of the
stack, and move it to the top of another stack, the working
stack. From there another forklift can move the boxes to a final
output stack. Physical limitations prevent boxes being moved from the
working stack to the input, or from the output to the working
stack. The desired outcome is that the output should be ordered with
box number 1 on top, then 2, then 3, \ldots, with box $n$ at the
bottom. An example of a sorting procedure in progress is shown in
Figure \ref{fork}. In this analogy, the working stack corresponds to
our forkstack data structure, and the operations performed on it by
the truck to its operations.

The process of sorting $236415$ is documented below. Note that, at the
stage shown in Figure \ref{fork} it is essential that 41 be moved as a
pair -- moving either 4 alone, or the triple 415 would result
(eventually) in 1 lying on top of 4 or 5 in the working stack, and
thereby prevent sorting.
\[
\begin{array}{r|r|rl}
\mbox{ Input } & \mbox{ Working } & \mbox{ Output } \\ \hline
236415 & & &\\
6415 & 23 && \\
415 & 623 & &\\
415 & 23 & 6&\mbox{ (See Figure \ref{fork})} \\
5 & 4123 & 6& \\
 & 54123 & 6& \\
 & 4123& 56 &\\
 & 123& 456&\\
 &   & 123456&\mbox{ Finished}
\end{array}
\]
Some permutations, such as 35142 cannot be sorted. Here, we may move 3
to the working stack, and then 5 to the output, but now whether we
move 1 alone, 14, or 142, we wind up with 1 lying on top of 3 or 4 in
the working stack, and cannot complete the sorting procedure. We will
see below that if we can avoid creating this type of obstruction in
the working stack, then sorting is possible.

\section{Definitions and formalities}

In the subsequent sections we will tend to continue to use the
terminology of the introduction speaking of the input stack,
forklifts, etc. However, it will be convenient to introduce a certain
amount of basic notation in order to facilitate discussion. As we will
always take the initial input to be a permutation of 1 through $n$ for
some $n$, the contents of each stack at any time can and will be
represented by sequences of natural numbers (not containing
repetitions). Our ultimate objective is always to reach a state where
the contents of the output stack are the permutation
\[
1 \, 2  \cdots (n-1) \, n
\]
and we will refer to this outcome as success.

In the basic situation where both forklifts are of unlimited capacity,
we use $\F$ to denote the collection of all permutations for which
success is possible. If the input to working stack forklift is limited
to moving $s$ boxes in a single move, and the working to output one to
moving $t$, then we denote the corresponding class $\F(s,t)$. Here $s$
and $t$ are either natural numbers, or $\infty$.

Given a permutation $\pi$ as input, a sequence of operations is {\em allowed},
if it does not result in an output state which provides clear evidence
that sorting is not being carried out. That is, a sequence of
operations is allowed if at the end of the sequence the output stack
contains some tail of $1 2\cdots n$.

In discussing the algorithms for sorting it will be helpful to
pretend that it is possible to move boxes directly from the input
stack to the output stack -- and such an operation, as well as the
more normal type of output is called {\em direct output}. So a direct
output move consists either of output from the working stack, or
moving a part of the input stack to the working stack (in a single
lift), and then moving exactly that set of boxes to the output stack,
again in a single lift.

When we consider enumeration results for forkstacks, it will often be
convenient to think in terms of which permutations of an original
input of $12 \cdots n$ can be produced, rather than which sequences
can be sorted. In that case it will be convenient to be speak of {\em
operation sequences}.  In the operation of a forkstack we use
$\sigma[k]$ to denote the operation of pushing $k$ elements onto the
stack, and $\tau[l]$ to denote the operation of popping $l$ elements
from it. If a parameter $k$ or $l$ is omitted, it is taken to equal
1.

\section{The sorting algorithms}

How should a fork stack actually carry out its task of sorting a
permutation when this is possible? It turns out that there is a
straightforward algorithm to accomplish this operation. Broadly
speaking, we may use a simple modification of a greedy algorithm:
\begin{itemize}
\item
perform any output as soon as possible,
\item
otherwise move the maximum decreasing sequence from the head of the
input onto the working stack.
\end{itemize}

In order to justify this claim (with some technical changes to the
second option) we require a slightly more abstract characterisation of
unsortability.

\begin{definition}
For positive integers $a$ and $b$, $a << b$ means $a < b -
1$. In a series of fork stack moves, we say that the {\em dreaded 13}
occurs if at some point the working stack contains adjacent elements
$ab$ with $a << b$.
\end{definition}

\begin{proposition}
\label{DREAD13}
A permutation $\pi$ is unsortable if and only if every allowable
sequence of fork stack operations that empties the input produces, at
some point, the dreaded 13.
\end{proposition}

\begin{proof}
Suppose that we cannot avoid producing a $13$.
Then we cannot sort $\pi$ for there is no way to insert the missing
elements into the gap between the elements $a << b$ witnessing the
$13$. On the other hand, if there is some allowable sequence of
operations that empties the input stack and avoids producing a 13,
then on completing them, the contents of the working stack will be a
{\em decreasing sequence, except possibly for some blocks of consecutive
increasing elements}. Such a stack is easily moved to the output in its
sorted order.
\end{proof}

We refer to a sequence of the type emphasised above, as a {\em
near-decreasing} sequence. Suppose that no immediate output is
possible and consider the maximal near-decreasing sequence, $\alpha$,
at the top of the input stack. If the symbols occurring in $\alpha$ do
not form an interval, then any move other than taking the whole
sequence $\alpha$ and transferring it to the top of the working stack
will, immediately or eventually, cause the dreaded 13. It will cause
an immediate 13 if we move more symbols than occur in $\alpha$, for
the transition between the final element of $\alpha$ and the next
element of the input is an increase of more than 1. If, on the other
hand, we break $\alpha$ at some intermediate point or points, then
eventually either the same 13 as above will be formed, or some symbol
of $\alpha$ from below a gap in its values will be placed directly on
top of some symbol from above that gap, thus creating a 13.  If the
symbols occurring in $\alpha$ do form a consecutive interval, then, as
above, they must still all be moved to the working stack before any
element of the remainder of the sequence is. However, we can arrange
to place them on the working stack in order, with largest
deepest. This is preferable to any other arrangement on the working
stack, for it makes the top element of the working stack as small as
possible, minimising the possibility of later creating a dreaded 13.

Doing direct output as soon as it becomes available can never
interfere with sorting. For if we have a successful sequence of
sorting moves which we modify by doing some direct output earlier, we
can simply continue to carry out the successful sequence, ignoring any
effect on symbols which have already been moved to output -- and we
will still succeed. So we may assume that any sorting algorithm does
in fact perform direct output whenever it can. Then the observations
of the preceding paragraph imply that when direct output is not
available, the maximal near-decreasing sequence at the top of the
input stack must be moved. If this sequence contains gaps, there is no
choice in how to move it, and we have argued that if it does not, then
moving it so that it forms an increasing sequence on the working stack
is at least as effective as any other choice.  This establishes that
Algorithm \ref{FORKSORT} will correctly sort any input stack, if it is
sortable at all.

\begin{algorithm}
\caption{Sorting with a powerful fork-lift}
\label{FORKSORT}
\begin{algorithmic}
\REPEAT
\STATE Perform as many direct output moves as possible.
\STATE{Move the maximal near-decreasing sequence from the top of the
input stack to the working stack, as a block if it contains gaps, so
that it becomes increasing if it does not.}
\UNTIL{input stack is empty}
\IF{working stack is empty}
\STATE Success!
\ELSE
\STATE Failure.
\ENDIF
\end{algorithmic}
\end{algorithm}

How does Algorithm \ref{FORKSORT} need to be modified in the case
where either or both of the forklifts moving from input to working
stack, or from working stack to output, are of limited power?  The
first issue is how to modify Proposition \ref{DREAD13}. The
13 configuration is bad regardless of the power of our forklifts, but
if our output lift is limited to moving $t$ boxes we must add the
condition that the working stack should not contain an increasing
sequence of length longer than $t$. Now modifying the algorithm is
straightforward. In the case where the maximal near-decreasing
sequence contains gaps it must be moved as a block to avoid 13's. So,
if this block is larger than the capacity of our working forklift, we
fail. In the non-gap case, we would normally attempt to make the
sequence increasing. Of course this would be foolish if it overwhelmed
the capacity of our output lift (and it could be impossible depending
on the capacity of our input lift). The only other choice that does
not create a 13 is to make it decreasing, so this should be attempted
if the first choice is unavailable. Failure may later occur because we
create a block that is too long to move in the working stack, or a 13
there, but if not, then the algorithm will succeed.

\section{Finite basis results}

We now begin our combinatorial investigation of the collections of
permutations sortable by various types of forklifts. The
problem which we address in this section is how to identify the
sortable or unsortable permutations without reference to Algorithm
\ref{FORKSORT}. In the following section we will consider the problem
of enumerating these classes. For identification purposes we
concentrate on producing a list of minimal unsortable permutations.

\begin{definition}
Given permutations $\sigma$ and $\pi$, we say that $\sigma$ is {\em
involved in} $\pi$, and write $\sigma \inv \pi$ if some subsequence
of $\pi$, of the same length as $\sigma$, consists of elements whose
relative order agrees with those of the corresponding elements of
$\sigma$. A collection of permutations closed downwards under $\inv$
is called a {\em closed class}.
\end{definition}

It is easy to see that each of the collections $\F(s,t)$ of sortable
permutations for a particular combination of forklifts is a closed
class. This is because we may sort any subsequence of a sortable
sequence by simply ignoring any moves that do not affect members of
the subsequence. This policy cannot increase the load on a forklift in
any single move, so it still sorts the remaining elements. It follows,
that if we take $U(s,t)$ to be the set of $\inv$-minimal unsortable
permutations then:
\[
\mbox{$\pi$ is $(s,t)$-unsortable}  \: \iff \: \mbox{$\sigma \inv \pi$ for
some $\sigma \in U(s,t)$.}
\]
In particular, $U(s,t)$ can be thought of as a description of
$\F(s,t)$ and we shall refer to it as the {\em basis} of
$\F(s,t)$. For example, the case $s = t = 1$ corresponds to sorting
with a single stack, and it is established in \cite{Knu} that
\[
U(1,1) = \set{213}
\]
This differs superficially from the cited result, owing to our
convention the output should be produced with largest deepest, so it
should be the {\em largest} input item which is popped first.

\begin{theorem}
For any $1 \leq s,t \leq \infty$ the set $U(s,t)$ is finite.
\end{theorem}

\begin{proof}
As in the case of the sorting algorithm, we will
first consider the case $s = t = \infty$, and then modify the
result to allow for the possibility of one or both forklifts being of
limited power.

In order to show that $U(\infty, \infty)$ is finite, it is sufficient
to establish that any unsortable permutation $\pi$ has an unsortable
subsequence $\sigma$ whose length is less than some fixed upper
bound. For in that case, the length of each element of $U(\infty,
\infty)$ is less than that upper bound, and of course there are only
finitely many such permutations. So, let an unsortable permutation
$\pi$ be given. As $\pi$ is unsortable, Algorithm \ref{FORKSORT} fails
to sort it, and so according to Proposition \ref{DREAD13}, it must at
some point produce a 13 in the working stack. We consider the state of
the system when the first move from input to storage which would
create a 13 is about to be made. The basic idea is that all of the
elements which contribute to the failure of the algorithm at this
point have a reason for being in the position that they are
in. The collection of these elements, together with the ones which
give them their reasons form the obstruction, $\sigma$, to sortability
whose size is bounded. We warn the reader that the actual execution of
this idea is of very limited interest, and if she is convinced of its
basic soundness it would probably be better to skip it.

Let the block from the input stack whose movement creates the first 13
be $B$ with top element $b$, the contents of the working stack just
prior to this move be $S$ with top element $s$, the remaining contents
of the input stack be $I$ with top element $i$, and the
contents of the output stack be $O$ with top element $o$ (if any). The
13 which the move creates is some pair $xs$ where $x$ is the bottom
element of $B$. 

As the algorithm specified making a block move from input, no direct
output can have been possible. In particular, since $s$ could not be
output directly, the largest remaining element $n_1$ smaller than $o$
must be different from $s$ (if $o$ does not exist, this element is
simply the largest element overall).  This element must be in $I$,
since were it in $S$ then we would either have a 13 already, contrary
to hypothesis, or would be able to output the block including it and
$s$, while were it in $B$, direct output from $B$ would be
possible. Note that an indirect consequence of this part of the
argument is that $I$ is non-empty.

Since the block $B$ was broken off the input between $x$ and $i$,
there must be an element $n_2$ with $x < n_2 < i$. This element might
belong to $B$, $I$, or $S$.

Since the stack is non-empty, the preceding move onto the stack moved
a block whose top was $s$ (or a block whose top was subsequently
output after some direct output from the input stack -- but such
elements are irrelevant). There are two possible ways in which this
block can have been broken off the input. Either it ended just above
the top of block $B$, or it ended just above some element which has
subsequently been output. In the first case, $s$ lies above some
element $n_3$ (which was part of its block), so that for some element
$n_4$ we have $n_3 < n_4 < b$. The element $n_4$ might be in $S$, $B$,
or $I$. In the second case, $s$ lay above the element $o$ in the
original input.

Finally, it might be necessary to ensure that the block $B$ is not a
block of consecutive elements (so that it is moved with $x$ at the
bottom, when rearrangement would avoid the 13). This would be
witnessed by the existence of an element $n_5$ with $x < n_5 < b$ and
$n_5$ in $S$ or $I$.

Now consider an attempt to sort the subpermutation of the original
permutation whose elements are:
\[
\set{b, x, s, o, i, n_1, n_2, n_3, n_4, n_5}.
\]
The elements $n_j$ (and in some cases, indirectly the other elements)
ensure that the $xs$ pair will be produced in the working stack, and
thus prevent sorting from taking place.

Next consider the class $\F(\infty, t)$ for some $t < \infty$. As the
operation of the input forklift is unrestricted, the only new
obstructions which might arise would occur when we had in the stack a
sequence of $t+1$ or more elements which were forced to be in
increasing order (read top to bottom), since these would have to be moved as
a block but couldn't be because of the output restriction. As we
are considering only new obstructions, we may take the $t+1$ largest
of those elements to form a consecutive block $a$ through
$a+t$. In the unlimited successful sorting of this sequence
they are placed on the stack in increasing order, the stack must
contain only smaller elements when they are added.  No direct output
affecting $a+t$ can occur while they are being added, so there must be
a larger element lying below them all in the input. If no direct
output is to take place while they are being added, then it is only
necessary that they not be in an order which would allow them to be
placed on the stack in decreasing order -- that is, in any decreasing
sequence of increasing blocks except
\[
a (a+1) \cdots (a+t) \quad \mbox{and} \quad (a+t) \cdots (a+1) a.
\]
In the former case, no interposing direct output can interfere with
placing them in the stack in decreasing order, so they are not forced
to be in increasing order on the stack, unless there is some
subsequent $b < a$ which must be placed in the stack before they are
removed. If so, this will also lengthen the block to be removed by 1
element, and so we can shorten the block by one element. 
In the latter case, this caveat also applies, but also any
interposing element which is to be output directly, would force the
block into ascending order. So the elements $a$ through $a+t$, the
element $b$, or an interposing element, if required, and a larger
element preventing direct output, are sufficient to ensure that $a$
through $a+t$ must be put on the stack in increasing order, and
therefore provide any potential new obstructions to sortability. For
example, one such new obstruction in the case $t = 2$ is:
\[
3 \, 2 \, 5 \, 1 \, 4.
\]

By running the sorting algorithm backwards we see that in general:
\[
\F(s,t) = \F(t,s)^{-1}.
\]
As the basis of the collection of inverses of elements of a class is
simply the collection of inverses of its basis, we can conclude that
the classes $\F(s, \infty)$ are also finitely based for any $s$. But
then the arguments of the preceding paragraph apply also to the class
$\F(s,t)$, and so all these classes are finitely based.
\end{proof}

Using the proof of the result above makes the computation of the sets
$U(s,t)$ relatively straightforward. 
The set $U(\infty, \infty)$ consists of the permutation $35142$,
together with 45 permutations of length six, and 6 of length
seven. The sets $U(1,t)$ are of particular interest in connection with
the next section and they are:
\begin{eqnarray*}
U(1,\infty) &=& \set{ 2314, 3124, 3142} \\
U(1, t) &=&  \set{ 2314, 3124, 3142, (t+1) \, t \, (t-1) \cdots 2 \, 1
\, (t+2)} \quad (t \geq 2) \\
U(1,1) &=& \set{213}.
\end{eqnarray*}

\section{Enumeration of $\F(1,t)$}

In the case where $s = 1$, that is, the push operation onto the stack
is restricted to moving a single element, there are no possible
choices in the sorting procedure. Pops from the storage stack must be
made whenever they are available, and pushes made otherwise. This
makes the enumeration of these classes relatively straightforward, at
least compared to the classes where more general pushes are
available, which we defer to the next section.

The basic plan is to search for structural requirements on sortable
permutations which are sufficient to develop algebraic relationships
that the generating function for the class must satisfy. We will find
the ordinary generating function of each class in this collection. As
in the case of finding bases for the class, it turns out that the
simplest instance to handle is the case $t = \infty$, and the
remaining instances can be derived from it by restriction in a fairly
obvious way. The generating function for this class will be denoted
$f_\infty$, and we use $x$ as the variable symbol.

Suppose then that we have some permutation $\pi \in \F(1, \infty)$. Choose $u$
to be the maximum integer such that the elements 1 through $u$ occur
in $\pi$ in decreasing order (thus, if 2 follows 1, $u = 1$). So
\[
\pi = \sigma_u \, u \, \sigma_{u-1} \, (u-1)  \cdots \sigma_2 \, 2 \,
\sigma_1 \, 1 \, \sigma_0
\]
for some sequences $\sigma_0$ through $\sigma_u$, where $u+1$ does
not occur in $\sigma_u$.

Consider now the sorting procedure. The elements of $\sigma_u$ are
processed, and then we come to $u$. Now by the choice of $u$, $u+1$
has not yet been processed, so we may not output $u$ (except in the
trivial case where all the $\sigma_i$ are empty). So $u$ must be moved
to the working stack. However, if it is non-empty at this time, that
move would create a 13. So the working stack must be empty, and
$\sigma_u$ must have been a sortable permutation of a final
subinterval of the values occurring in $\pi$. Now proceed to the stage
where $u-1$ is about to be moved. Again, either $u+1$ has turned up by
now, and the working stack is empty, or it contains only the value
$u$. In either case $\sigma_{u-1}$ is a sortable permutation of a
final subinterval of the remaining values. This argument persists
inductively. So in the end we see that $u+1$ occurs in the first
non-empty $\sigma_j$, and that the general requirements for
sortability are that $\sigma_i$ be sortable for each $i$, and that
each $\sigma_i$ be supported by an interval, with $\sigma_0 < \sigma_1
< \cdots < \sigma_u$.  

In other words, having determined $u$, we are free only to decide the
sizes of the individual $\sigma_i$, and then their structure within
the class, but having chosen their sizes, the elements that they
contain are fixed. To carry out the enumeration, we distinguish two
cases according to whether or not $\sigma_0$ is empty. If it is, then
$\pi = \pi' 1$ where $\pi'$ is an arbitrary sortable
permutation. Permutations of this type are enumerated by $x
f_{\infty}$. If $\sigma_0$ is not empty, then the generating function
for the collection of permutations $\pi$ of this type (with $u$ fixed)
is
\[
(x f_\infty)^u (f_{\infty} - 1).
\]
We can sum this over the possible values of $u$, and include the
trivial case of an empty permutation to obtain the equation:
\[
f_\infty = 1 + x f_{\infty} + (f_{\infty}-1) \sum_{u=1}^{\infty} x^u f_\infty^{u}
\]
or, after summing the geometric series:
\[
f_\infty = 1 + \frac{x f_{\infty}^2 - x^2 f_{\infty}^2}{1 - x f_{\infty}}.
\]
We can then solve the resulting quadratic to get:
\[
f_{\infty} = \frac{1 + x - \sqrt{1 - 6 x + 5 x^2}}{4 x - 2 x^2}
= \frac{2}{1 + x + \sqrt{1 - 6 x + 5 x^2}}.
\]
The sequence that this generating function defines:
\[
1,\, 1\,, 2,\, 6,\, 21,\, 79,\, 311,\, 1265,\, 5275 \ldots
\]
is number A033321 in \cite{Slo}, and the references provided for it there
connect it with other interesting enumeration problems.

The only change that needs to be made to find the generating function
$f_t$ for $\F(1,t)$ is to change the upper limit of summation in the
relationship above from $\infty$ to $t$, since the maximum increasing
sequence that we can deal with on the working stack is of length
$t$. Of course $f_1$ is the generating function for the Catalan
numbers. In general, however, this gives an algebraic equation
satisfied by $f_t$ with coefficients that are polynomials in $x$. The form of
this equation is slightly simpler for the related function $g_t = x
f_t$ namely
\[
g_t^{t+1} + (1-x) ( g_t^t + g_t^{t-1} + \cdots + g_t^{2} ) - g + x = 0.
\]
which can be simplified still further through multiplication
by $g-1$ yielding
\begin{eqnarray*}
0 &=& g_t^{t+2} - x g_t^{t+1} + (x-2) g_t^2 + (x+1) g_t - x \\
{} &=& x ( - g_t^{t+1} + g_t^2 + g_t - 1) + (g_t^{t+2} - 2 g_t^2 + g_t).
\end{eqnarray*}

This allows efficient exact enumeration of these classes
using standard generating function techniques. It also allows
asymptotic expansions of the form:
\[
c_n = \frac{r^{-n}}{n^{3/2}} \left( \sum_{k=0} \frac{e_k}{n^k} \right)
\]
to be computed to any desired degree of accuracy using the methods
developed in \cite{FS}.

The behaviour of the radius of convergence $r$ (whose reciprocal gives
the exponential part of the growth rate for the coefficients $c_n$),
as $t$ increases from $1$ to $\infty$ is particularly interesting. It
begins at $1/4$, since $t = 1$ gives us the Catalan numbers and then
decreases to $1/5$ at $t = \infty$. However, the rate of convergence
to $1/5$ is geometric, with the difference decreasing by roughly a factor
of 3 at each step. The first six values are:
\[
.2500, \, .2114, \, .2033 , \, .2010 , \, .2003 , \, .2001
\]
We can justify these results formally by noting that the radius of
convergence in each case is the smallest positive root of the
discriminant of the polynomial which $g_t$ satisfies (this follows
easily from results in \cite{FS}).  Finding this root is simplified by
considering the second form of the equation for $g_t$, namely (after
change of name for convenience):
\[
x (- a^{t+1} + a^2 + a - 1) + (a^{t+2} - 2 a^2 + a) = 0.
\]
We seek a minimum positive value $x_0$ of $x$ for which the resulting
polynomial in $a$ has a double root. In order for this to be true,
that root will also be a root of the derivative with respect to $a$ of
this equation, that is of
\[
x ( -(t+1) a^t + 2 a + 1) + ((t+2) a^{t+1} - 4 a + 1) = 0.
\]
Eliminating $x$ between these two equations gives:
\[
a^{t+1} ( a^{t+1} - t a^2 + (t-3) a + 2)  - 3 a^2 + 4 a - 1 = 0.
\]
This equation has a root near $a = 1/3$ which corresponds to the
smallest root $x$ that we seek. If we write that root in the form
$a = 1/3 + e_t$, then the final three terms will be of order $2 e_t$,
while the value of the first expression will be of order
$(1/3)^{t+1}$. In other words, the value of $e_t$ decreases
geometrically to $0$ as $t \to \infty$. Then substituting this value
of $a$ and solving for $x$ shows that the difference $x - 1/5$ is also
geometrically decreasing. 

\section{Enumeration for $\F(s,t)$ ($s,t > 1$)}

In carrying out the enumeration of $\F(1,t)$ we concentrated on the
structure of the sortable permutations. One reason for doing this was
that, even in this simple context, the series of operations required
to sort a permutation is not uniquely defined. For example $21$ can be
sorted either by a sequence of alternating pops and pushes, or by two
single pushes, followed by a pop of the whole storage stack. When both
$s$ and $t$ are larger than 1, this ambiguity in the operation
sequence is compounded in a very complex fashion, and as we shall see,
makes it difficult to carry out explicit enumeration. In fact, the
only class of this sort for which we will provide an explicit
enumeration result is when $s = t = 2$.

In this section we will consider instead of the class of sortable
permutations, the class of permutations which we can generate,
i.e. produce by some sequence of operations from initial sorted
input. As this class consists simply of the inverses of the sortable
permutations, there is no essential difference involved in considering
it instead. We shall be concentrating on operation sequences, and the
relationship, $\equiv$, of equivalence between them defined as
producing the same output from initially sorted input. The goal of the
argument then is to produce a representative for each equivalence
class of operation sequences. We will show that it is possible to
define a deterministic push-down automaton that recognizes precisely
one operation sequence from each equivalence class.

Since, in particular, the language accepted by such an automaton is an
unambiguous context free language, the results of section 2 in
\cite{ChS} imply that the multivariate generating function for a
representative class of operation sequences satisfies an algebraic
equation. By replacing the (variables corresponding to) a push of $i$
items by $x^i$, and any pop operations by $1$, each term which
represents a permutation of length $n$ is replaced by $x^n$ and so we
obtain the generating function for permutations which we can generate.
Thus we may conclude that the generating functions for these
classes of permutations are algebraic.

At present, the only class for which we have carried out the details
of this construction explicitly is when $s = t = 2$. We conclude the
section with a summary of the results for that case.

The structure of the argument is as follows:
\begin{itemize}
\item
First, we introduce a simple form of reduction for operation
sequences, which allows us to consider only operation sequences which
are in reduced form.
\item
Second we argue that if two reduced operation sequences
represent the same permutation, then they must have the same
``profile'' (as defined below).
\item
Finally we argue that even when two reduced operations sequences
have the same profile, they can only represent the same permutation in
certain special circumstances, circumstances so restrictive that, when
the operation sizes are bounded we
can find a suitable automaton.
\end{itemize}
It will be noted that as usual we deal primarily with the case $s = t
= \infty$, introducing the operation bounds only at the last moment,
in each phase of the argument.

In the operation of a forkstack let $\sigma[a]$ denote the operation
of pushing $a$ elements onto the stack, and $\tau[b]$ the operation of
popping $b$ elements from it. If the parameter $a$ or $b$ is omitted,
it is taken to equal 1. A well formed operation sequence is then a
word in the symbols $\sigma[a]$ and $\tau[b]$ with $a$ and $b$ running
over the positive integers, which has the property that for any
initial segment, the sum of the push sizes is at least as great as the
sum of the pop sizes, and over the whole sequence those sums are
equal. The size of an operation sequence is the sum of all its push
(or pop) sizes, that is, it is the size of the input sequence which it
rearranges. Two operation sequences are equivalent if they have the
same size, $n$, and produce the same permutation when acting on input
$12\cdots n$.

\begin{figure}[ht]
\begin{center}
\epsfysize=4cm
\epsffile{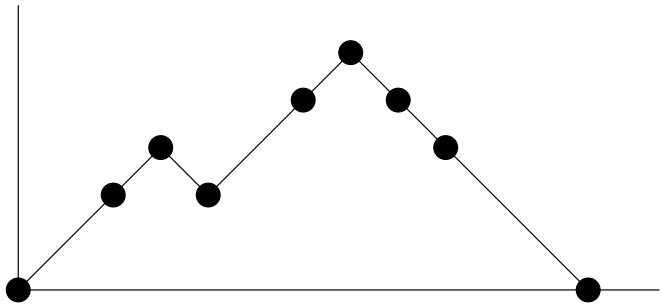}
\caption{A lattice path for producing  $512463$}
\label{lattpath}
\end{center}
\end{figure}

We associate with each operation sequence a series of vertices in the
upper quadrant of the plane. If the sequence is
\[
\alpha_1 \alpha_2 \cdots \alpha_m
\]
then the associated vertices are $v_0$ through $v_m$, where $v_0 =
(0,0)$ and for $i \geq 1$:
\[
v_i =
\left\{
\begin{array}{ll}
v_{i-1} + (a_i,a_i) & \mbox{if $\alpha_i = \sigma[a_i]$,} \\
v_{i-1} + (a_i, -a_i) &  \mbox{if $\alpha_i = \tau[a_i]$.}
\end{array}
\right.
\]
We also associate with this sequence the path formed by the union of
the closed line segments $[v_{i-1}, v_i]$ for $1 \leq i \leq m$. This
path, which we call the {\em profile} of the operation sequence, is of
course a Dyck path and, in the restricted case where only single pops
and pushes occur, the fact that distinct paths represent inequivalent
operation sequences is one of the methods for connecting
stack-sortable permutations to the Catalan sequence. Much of the
difficulty in dealing with more general operation sequences arises
because the relation of equivalence is not directly connected either
to paths, or to their vertices. In figure \ref{lattpath} we illustrate
a lattice path for producing the sequence $512463$. Notice that
because the output occurs onto a stack, the first element popped from
the working stack actually becomes the last element of the output permutation.

Suppose that within an operation sequence we have a consecutive pair
of elements $\sigma[a] \tau[b]$ with either $a > 1$, or $b > 1$. If $a
\geq b$ we may replace this pair by the sequence $\sigma^b \tau^b
\sigma[a-b]$, and thus produce an equivalent sequence. 
If $a < b$ we may replace it with with $\tau[b-a] \sigma^a
\tau^a$. By performing a sequence of such reductions we can produce an
equivalent operation sequences in which the only peaks in the profile
are represented as pairs $\sigma \tau$. At that point, no further
transformations of this type are available. We will call an operation
sequence with this property, {\em reduced}. We have seen that
every operation sequence is equivalent to at least one reduced
one (``at least'' is necessary here since for example $\sigma[3]
\tau[3]$ is equivalent to both of $\sigma[2] \sigma \tau
\tau[2]$ and $\sigma^3 \tau^3$). 

Consider the form of a permutation produced through the operation of a
forkstack according to a reduced operation sequence,
\[
\alpha = \alpha_1 \alpha_2 \cdots \alpha_m. 
\]
At each peak, a sequence of pushes, culminating in the push of a
single item, is followed by a sequence of pops, beginning with the pop
of a single item. Of course this item, call it $p$, was the same as
the one pushed by the final push -- we say that $p$ is produced by
this particular $\sigma \tau$ pair. As all the elements larger than $p$
are still in the input at the time that it is popped we see that, in
the final permutation $\pi$ produced by the sequence, $p$ will be
larger than all the elements that follow it. Such an element will be
called a local maximum of $\pi$ read right to left. All
other pops which directly follow this particular peak (i.e. which take
place before any more pushes) will produce
elements in $\pi$ which were already in the stack before $p$ was
added, that is, elements smaller than $p$. This argument establishes:

\begin{lemma}
Let $\alpha$ be a reduced operation sequence which produces a
permutation $\pi$. The elements of $\pi$ which are produced by the
$\sigma \tau$ pairs forming the peaks of the profile of $\alpha$ are
the local maxima of $\pi$ read right to left.
\end{lemma}

The main reason for proving this lemma is to make use of it in
showing that the profile of a reduced operation sequence is an
invariant of its equivalence class. That is:

\begin{theorem}
Any two equivalent reduced operation sequences have the same
profile.
\end{theorem}

\begin{proof}
Suppose that $\alpha$, as above, is a reduced sequence producing the
permutation $\pi$. By the preceding lemma, we can identify the
elements of $\pi$ which arise from peaks in $\alpha$. Any two peaks
are separated in the profile by a descent followed by a rise. The
total length of the descent is the number of elements popped between
the peaks, and so is equal to the size of the gap between the two
corresponding elements of $\pi$. The size of the rise is equal to the
difference in the values of the local maxima. As these quantities
depend only on $\pi$ and not on $\alpha$, the profile of $\alpha$ is
determined by $\pi$.
\end{proof}

Next we will show that some of the segments making up the profile of a
reduced sequence $\alpha$ must also occur in any reduced
sequence $\beta$ equivalent to $\alpha$. Unimaginatively, we refer to
these as {\em fixed operations}. In order to characterise the fixed
operations we need to introduce the notion of {\em corresponding
operations} within an operation sequence. These are most easily
understood in terms of the forkstack operation. Namely, we define the
{\em support} of each push operation to be the set of elements that it
places on the stack (which is, of course, an interval), and of each
pop operation to be the set of elements which it removes from the
stack (not necessarily an interval). A push and a pop
operation form a corresponding pair if their supports have non-empty
intersection. Graphically, a push and a pop operation form a
corresponding pair if there is a horizontal line segment joining some interior
point of the segment of the profile corresponding to each of the
operations which intersects the profile only at its endpoints (a
connected piece of a contour line at a non-integer height).

\begin{proposition}
If the subsequence of pops corresponding to a particular push, in an
operation sequence is not a subword of the operation sequence, then
that push is a fixed operation.
\end{proposition}

\begin{proof}
Let $\alpha$ be an operation sequence, and $\lambda$ an element of
$\alpha$ representing a push that satisfies the conditions of the
proposition. Certainly the size, $a$, of $\lambda$ is at least two.
Suppose that $\lambda$ pushes elements $x+1$ through $x+a$ onto the
stack. Up to and including the first pop, $\nu$, corresponding to
$\lambda$, the output will consist of some elements smaller than $x+1$
(output before the push $\lambda$ was made), a non-empty interval
$[x+a+1, b]$, and an interval $[x+1, x+u]$, for some $u < a$ ($u$ will
be the length of the overlap between the segment of the profile
associated with $\nu$, and that associated with $\lambda$). Among the
subset of these elements which are larger than $x$, the first element
output (corresponding to a peak) will be larger than $x+a$. 

Suppose that $x$ had not been output before the push $\lambda$ was
made. Then before $x+a$ or $x$ are output, some other element $c > b$
representing any one of the peaks separating the pops corresponding to
$\lambda$ will be output. So, from the permutation alone we can
determine that $x+1$ was added to the stack above both $x$ and $x+a$
(but not $x+a+1$). That is, the push of the interval $[x+1, x+a]$ is
fixed, as claimed.

On the other hand, if $x$ had already been output when the push
$\lambda$ was made, then it is separated from $x+1$ in the output by
an element larger than $x+a$, and using the remainder of the argument
in the previous paragraph we can again deduce that $x+1$ was
added to the stack above both $x$ and $x+a$ (but not $x+a+1$), and
hence that the push $\lambda$ is fixed.
\end{proof}

Next we can find some fixed pops:

\begin{proposition}
If a push in an operation sequence is
fixed then each of the pops that correspond to it is also fixed.
\end{proposition}

\begin{proof}
Let $\lambda$ be such a push, and $\nu$ such a pop. First consider the
case where $\nu$ pops only elements which were pushed by
$\lambda$. Then $\nu$ partitions the support of $\lambda$ into three
pieces, an initial segment $I$, a segment $J$ which is the support of
$\nu$, and a remainder $K$. In the output stack, all the elements of
$I$ are lower than all the elements of $J$  which are
in turn lower than all the elements of $K$. Since $I$, 
$J$, and $K$ occurred in that order in the input stack (as $\lambda$
was fixed), all of the elements in $I$ had to be popped before any
elements of $J$, and all of them before any elements
of $K$. Furthermore the elements of $J$ must have
been popped as a block since they occur in order in the output
stack. That is, $\nu$ is fixed. Essentially the same argument applies
when the segment representing $\nu$ extends above, below, or both
above and below, the ends of the segment representing
$\lambda$. Consider, for the sake of illustration, the first of these
possibilities. Then $\nu$ pops from storage a block consisting of
elements pushed by operations later than $\lambda$, and then a non-empty
initial segment $I$ of the block pushed by $\lambda$. If the initial
part of this segment were broken up in any way (or extended further)
then elements of the output which should be above $I$ would be below
it, or vice versa. If the end of $\nu$ were pushed downwards, then
elements not in $I$ would finish below it in output, again an
inconsistency with the output permutation.
\end{proof}

By repeating the arguments above, or more simply by noting that we
could read an operation sequence in reverse, exchanging pops and
pushes we obtain:

\begin{theorem}
Let $\alpha$ be a reduced operation sequence. Any operation
whose corresponding operations do not form a subword of $\alpha$ is
fixed, as is any operation corresponding to a fixed operation.
\end{theorem}

\begin{definition}
A {\em rise-fall} subsequence of an operation sequence is
a minimal pair of subwords, the first of which consists entirely of
pushes, and the second entirely of the corresponding pops, both of the
same size. Two operation sequences are {\em rise-fall} related, if one is
obtained from the other by replacing one rise-fall subsequence with
another (in the corresponding part of the operation sequence)
without changing the output permutation.
\end{definition}

For example, in the operation sequence:
\[
\sigma[3]^a \sigma[2] \sigma[1]^b \tau[1]^b \tau[1] \sigma[1]^c
\tau[1]^c \tau[1] \tau[2]^a \tau[1]^a
\]
there are three rise-fall subsequences, $a$, $b$, and $c$ (the latter
two in some sense trivial) as marked by the superscripts. Write this
sequence as $\sigma[3] \beta \tau[2] \tau[1]$. It is rise-fall related
to both of:
\[
\begin{array}{l}
\sigma [1] \sigma [2] \: \beta \: \tau [1] \tau [2] \quad \mbox{and,} \\
\sigma[2] \sigma[1]  \: \beta \: \tau[3],
\end{array}
\]
because
\[
\sigma[3] \tau[2] \tau[1] \equiv \sigma[1] \sigma[2] \tau[1] \tau[2]
\equiv \sigma[2] \sigma[1] \tau[3].
\]

\begin{theorem}
Equivalence for reduced operation sequences is the transitive
closure of rise-fall relatedness.
\end{theorem}

\begin{proof}
Almost by definition, the only changes that can be made to a reduced
operation sequence in order to form another such sequence modify
unfixed pushes or pops. In a reduced operation sequence
$\alpha$, consider any maximal subword $\alpha_i \alpha_{i+1} \cdots
\alpha_m$ of unfixed pushes. If the last pop corresponding to
$\alpha_i$ ended strictly below $\alpha_i$ it would either correspond
also to a fixed operation, $\alpha_{i-1}$, or in the event that
$\alpha_{i-1}$ were a pop, its corresponding operations would not form
a subword of $\alpha$. So, it would be fixed, and hence so would
$\alpha_i$ be. Thus the last pop $\alpha_n$, corresponding to
$\alpha_i$ ends at the level where $\alpha_i$ begins. Similarly, the
first pop, $\alpha_j$, corresponding to $\alpha_m$ begins where
$\alpha_m$ ends. Either the subword from $\alpha_j$ through $\alpha_n$
consists of exactly the pops corresponding to elements $\alpha_i$
through $\alpha_m$, or there are some interposed valleys and peaks. In
the latter case, by the same argument as above, the bottoms of these
valleys must occur at junction points between the elements $\alpha_i$
through $\alpha_m$, and these junction points are determined by the
positions of the various elements pushed by $\alpha_i$ through
$\alpha_m$ relative to the local maxima of the output permutation read
right to left. In other words, $\alpha_i \alpha_{i+1} \cdots \alpha_m$
breaks apart into a sequence of the rise parts of rise-fall
subsequences. The elements on which these rises operate are determined
by the profile, and so the only scope for modification in producing an
equivalent sequence is in changing some or all of these rise-fall
subsequences through rise-fall equivalences. These changes operate
independently of one another, and so we simply obtain the transitive
closure of rise-fall relatedness.
\end{proof}

It remains to characterize rise-fall relatedness, that is to determine
which sequences of the form:
\[
\sigma[a_1] \sigma[a_2] \cdots \sigma[a_k] \tau[b_l] \tau[b_{l-1}]
\cdots \tau[b_1]
\]
represent the same permutations.

By the condition on minimality:
\[
a_1 + a_2 + \cdots + a_i = b_1 + b_2 + \cdots + b_j
\]
if and only if $i = k$ and $j = l$. 

Let $m = a_1 + a_2 + \cdots + a_k$, and let:
\begin{eqnarray*}
s_i &=& a_1 + a_2 + \cdots + a_i \\
t_j &=& b_1 + b_2 + \cdots + b_j.
\end{eqnarray*}
In order to analyse which other sequences this might be rise-fall
related to we return to the analogy of reordering a stack of boxes
using two forklifts. When considering rise-fall sequences the
operations consist of a series of moves from the input stack to a
storage stack, followed by another series of moves from the storage
stack to the final output. So we represent the original input sequence
$1 \, 2 \, \cdots \, m$ as a column of numbers, with $1$ at the top.
We {\em mark} this column (between elements) on the left hand side
after the first $s_i$ elements for each $i < k$, and on the right hand
side after the first $t_j$ elements for each $j < l$. If we think of
the markings as slips of paper, say red paper for the left marks, and
blue ones for the right marks, then they can be used to instruct the
forklift operators how to carry out the rise-fall sequence. The pushes
are done by breaking the input stack at each red slip. Then the pops
are carried out by breaking the storage stack at the blue slips. The
minimality criterion is precisely that there is no pair of boxes
having both a red and a blue slip between them.

Obviously the blocks of the input between the marks are left intact,
but some rearrangement of those blocks is carried out. If we could
argue that the blocks originally consecutive in order were not
consecutive and in the same order in the output, and that we could
determine the markings from the rearrangements of the blocks alone,
then it would follow that two distinct sequences represent different
permutations. Of course we cannot prove this, for we have already seen
that equivalence is a non-trivial relation on rise-fall sequences.
However, we shall show that all such sequences which belong to
non-singleton equivalence classes belong to two well-defined types:
$k = l = 2$, and $k = 1$ or $l = 1$.

To do this, replace each block by a single element. There are now $n = k+l-1$
elements, $k-1$ markings on the left, and $l-1$ on the right. We
represent the marks as a sequence of $L$'s and $R$'s, representing the
side on which each mark occurs, reading from the top of the input
stack downwards. We will argue that, provided $n > 3$ and there are
marks on both sides prior to the last one, then we can determine {\em
from the output permutation alone} the
type ($L$ or $R$) of the final mark, and the permutation which arises from
deleting it and the last element.

Suppose first that the final mark is on the right. The last push move
will then be some sequence from $j$ through $n$. This will be followed
by singleton pops, leaving the tail of the output permutation as
$(n-1) (n-2) \cdots j$.  Assuming that this suffices to identify the
position of the final mark, we can also identify the form of the
permutation obtained by deleting $n$ and the final mark. It is
obtained by deleting $n-1$ from the output permutation and changing
$n$ to $n-1$. Note also in this case that $n$ never immediately
follows $n-1$ in the output. Now suppose that the final mark were on
the left. This time, the top of the storage stack, after all the
pushes have been made, begins with $n$ followed either by a descending
sequence beginning with $n-1$ (corresponding to other single pushes),
or by a block from $j$ through $n-1$ for some $j$. In the first case,
the entire descending block plus one more element is removed by the
first pop, leaving an output tail of the form $n (n-1) (n-2) \cdots
(n-k) r$ where $r < n-k-1$. Here note that the tail definitely differs
from the right-marked case. In the other situation we begin with a
doubleton pop leaving a tail of $n j$, again different from the
right-marked case. In either case simply deleting $n$ from the final
permutation leaves the permutation that results from deleting it from
the input also. So, these cases can be identified from the form of the
output permutation.

If $n > 3$ and all the marks save the last are on the same side,
then we can tell if the final mark is on the opposite side and also
which side the original marks were on. For note that the permutation
marked by $LL \cdots LR$ is $n \, (n-2) \, (n-3) \cdots 2 \, 1 \,
(n-1)$, that marked by $RR \cdots RL$ is $(n-1) \, (n-2) \, (n-3)
\cdots 2 \, 1 \, n$. If the final mark on the same side, then it is
still not possible to determine the marking as both $LL \cdots L$ and $RR
\cdots R$ produce $n \, (n-1) \, (n-2) \cdots 2 \, 1$.

We see that our goal of identifying the final mark, and the
permutation arising when the final element is deleted is achievable if
$n > 3$ and if marks occur on both sides. The only ambiguities that
might seem to arise in the entire mark sequence then are either in the
first two positions, or in a block of left marks at the top which
might be changed to right marks. However, it is easily checked that
the latter case results in a different output permutation provided
that there are marks on both sides in the whole
permutation. Furthermore, the only case where the output permutation
contains consecutive elements $i$ and $i+1$ in that order occurs when
$n = 3$, where we can produce $312$ through a $LR$ marking, and $231$
through a $RL$ one. Adding an additional mark on either side in either
case breaks the consecutive pair, and we have seen above, that no
later consecutive pairs can be formed.

What ambiguity does that leave us at the original level of
elements rather than blocks? If no pair of blocks which were originally
consecutive in the input are consecutive in that order in the output 
sequence then they are uniquely identified. So, if there are at least
three marks and they occur on both sides, then the final
permutation determines the marking (that is, the operation
sequence) completely. If the blocks are descending, then all the
marks go on one side, but we are free to choose which, and this
gives two operation sequences producing a single permutation. These
two sequences are represented in general by:
\[
\sigma[c_1] \sigma[c_2] \cdots \sigma[c_k] \tau[m] \quad \mbox{and}
\quad \sigma[m] \tau[c_1] \tau[c_2] \cdots \tau[c_k]
\]
for any sequence $c_1, c_2, \ldots, c_k$ of sum $m$.  In subsequent
arguments we will always assume that the first of these sequences is
chosen as the representative of its equivalence class, in situations
where both are available (either one might overwhelm the capacity of
one of our forklifts). This assumption is purely a matter of convenience.

If there are only two marks, one on either side, then fixing one, we
are free to move the other arbitrarily (or to remove it if we
wish). That gives for each pair $(j,m)$ with $1 \leq j < m$, a family
of $m$ equivalent operation sequences all producing the permutation
\[
(j+1) \, (j+2) \, \cdots m \: 1 \, 2 \cdots j.
\]
These sequences are:
\[
\begin{array}{c}
\sigma[1] \sigma[m-1] \tau[j-1] \tau[m-j+1] \\
\sigma[2] \sigma[m-2] \tau[j-2] \tau[m-j+2] \\
\vdots \\
\sigma[j] \sigma[m-j] \tau[m] \\
\sigma[j+1] \sigma[m-j-1] \tau[m-1] \tau[1] \\
\sigma[j+2] \sigma[m-j-2] \tau[m-2] \tau[2] \\
\vdots \\
\sigma[m] \tau[j] \tau[m-j].
\end{array}
\]
Finally, there is the case where there are no marks, arising from an
original pair $\sigma[m] \tau[m]$. This of course represents the
identity permutation, but that permutation can also be constructed in
many other ways by allowing left and right marks, all at the same
levels. We will always assume that the chosen representative for this
equivalence class is $\sigma^m \tau^m$.

In the case where there are no limits on the sizes of the push and pop
operations, each other non-trivial equivalence class of rise-fall
sequences contains exactly one element which involves a series of
pushes followed by a single pop. 

We intend to show that it is possible to construct a deterministic
push down automata which recognizes only fully reduced sequences, and
exactly one such from each equivalence class of operation sequences
when the size of a push is bounded by $s$ and that of a pop by
$t$. Once again, the construction will be more transparent if we first
ignore the restriction on the sizes (though of course this requires us
to allow infinitely many states), and then to show that introducing
those restrictions trims the automaton down to a manageable, that is
finite, size.

\begin{theorem}
\label{DPDA}
Let $1 \leq s, t < \infty$ be given. There exists a deterministic push
down automata whose accepted language consists only of fully reduced
operation sequences for the bounded forkstack where pushes up to size
$s$ and pops up to size $t$ are allowed, and which contains exactly
one such sequence from each equivalence class.
\end{theorem}

\begin{proof}
We begin by constructing a rudimentary automaton to recognize any
reduced operation sequence without bounds on the sizes of push or pop
operations. This automaton simply has a stack holding two types of
symbols, which we will call red plates and white plates, together with
a register which records whether the preceding operation was a push or
a pop. In the initial state the stack is empty. Each push operation
$\sigma[j]$, beginning from a state where the top plate is red (or the
empty stack), simply puts $j-1$ white plates, and one red plate onto
the stack, and records the last operation as a push. A pop operation
$\tau[l]$ basically pops the top $l$ elements from the stack. However,
it can fail in two ways: if the stack contains fewer than $l$ plates
making the pop impossible, and also if the preceding operation was a
push, and $l > 1$ or $l = 1$ and after the pop the top plate of the
stack is not red. The accepting states are any state with an empty
stack. It is clear that this automaton recognizes precisely the
reduced operation sequences.

In order to recognize only a single element from each equivalence
class we augment the automaton with green and amber plates, a
dictionary, and a notepad. The dictionary contains a
chosen representative of each equivalence class of rise-fall
sequences. Metaphorically, the purpose of the notepad is to allow us
to keep a record when we are making pops which might possibly be part
of a rise-fall sequence. If we discover this to be false, we stop
making notes, and return to a blank page. If we determine that such a
sequence has in fact occurred, we check it against the dictionary to
see whether it is the chosen representative of its particular
equivalence class. The purpose of the new coloured plates are to
provide markers where rise-fall sequences are broken up by intervening
operations. When a push operation follows a pop, before adding plates
to the stack the top plate is replaced with a green plate if it was
white, and an amber one if it was red. Green plates definitely break
up any potential rise-fall sequences and so will cause recording on
the notepad to cease. Amber plates behave as green plates if they
occur internally within a potential rise-fall sequence, but as red
ones if they form either end of such a sequence.

Formally then, we begin recording on the notepad each time that a pop
sequence begins with a red or amber plate on top of the stack. If at
any point during a recording sequence we pop a green or amber plate,
we discard the current contents of the notepad, and do not begin
recording again until we reach a state where the top of the stack is
red or amber. On the notepad we record the size of the pop made, and
the size(s) of the pushes that it matches. If, after making the pop,
the top of the stack is again red or amber, we check to see whether we
have matched exactly one push (in which case we fail if the pop we
just made was of more than one element), two pushes $a$ and $b$ or some
sequence of $k \geq 3$ pushes. If there were two pushes we determine whether
\[
\sigma[a] \sigma[b] \tau[a+b] 
\]
is the chosen representative of that equivalence class. If so, we
discard the notes, and begin a fresh set. If not, we fail. If we
matched at least three pushes, then we simply continue with a fresh
set of notes.

If, on the other hand, after the pop the top is white we check to see
whether we have already consumed at least two pushes. If so, we
discard the notes and wait until the next time the top of the stack is
red or amber before beginning to record again. If not, but the next
element is a push, we again discard the notes. If the next element is
a pop, we do it, and then check whether the two pops we have made have
consumed exactly one or two pushes. If so, we check this sequence
against the dictionary to see whether it is the chosen representative,
and continue if it is, discarding these notes as usual, failing
otherwise. If our two pops have not yet used up a single push, we
continue with pops (discarding notes if there are any intervening
pushes as usual) until we have either overrun, or precisely matched a
single push. In the latter case we fail, since our chosen
representative of that type consisted of a series of pushes followed
by a single pop.

Having made all these modifications it is now clear that our machine
recognizes exactly the reduced sequences all of whose rise-fall
subsequences are the chosen representatives of their equivalence
classes, and only these.

Now consider the case where the size of a push is limited to $s$ and
that of a pop to $t$. A minor modification is required in the situation
where a series of pops matches a single push. In the limited power
situation we now fail only if each of the pops could have been handled
as a push and the whole sequence of elements could have been handled
with a single pop. Otherwise, we discard notes and continue fresh.

The only situation in which we need to record sequences of more than
four moves on the notepad is when a series of pops are being matched
to a single push. This can lead to failure only if the total size of
the pops is at most $t$ (the pop limit) and so the length of such a
sequence is certainly bounded by $t+1$ (we need to record the push as
well).  In any case, the notepad, and hence the automaton, has only
finitely many states. Of course the alphabet is finite as well,
consisting of $s$ symbols for pushes, and $t$ for pops. The resulting
automaton still recognizes precisely a single representative of each
equivalence class of fully reduced operation sequences in the reduced
machine, exactly as we required.
\end{proof}

\begin{corollary}
For any natural numbers $s$ and $t$, the class of permutations which can be generated by a forkstack whose
pushes are limited to size at most $s$, and whose pops are limited to
size at most $t$ has an algebraic generating function. By reversal,
this is also true of the class of permutations sortable by such a
machine.
\end{corollary}

For the case where $s = t = 2$ we have carried out the construction of
the deterministic push down automaton, analysed the resulting
unambiguous context free grammar, and applied the methods of
\cite{ChS}\footnote{We reassure any properly sceptical reader
that each part of this computation, as well as the final
generating function, was checked in at least two ways to be correct
for all permutations in the class and their corresponding operation
sequences up to length 12.} in order to determine the
generating function $f$ of this class. The algebraic equation that $f$
satisfies is:
\[
x^4 f^4 - (x^4 + 2 x^3 - x^2) f^3 + (x^3 + 2 x^2 - 3 x) f^2 - (x^2 - 2
x - 1) f - 1 = 0.
\]
The radius of convergence of $f$ is the least positive root of:
\[
3 x^8 + 8 x^7 + 52 x^6 + 136 x^5 + 282 x^4 - 264 x^3 + 228 x^2 - 8 x -
5 = 0
\]
and has value approximately $0.18476$. Note that this is already
smaller than the radius of convergence in the case $s = 1$, $t =
\infty$ which was $0.2$. This corresponds to a larger exponential rate
of growth in the class size. In other words, it is better to have two
forklifts each of which can move two boxes, than one which can move
only one, regardless of how powerful the other one is!

In fact, we can extend Theorem \ref{DPDA}  to the case where one or the other
(but not both) of $s$ or $t$ is infinite. Consider the latter case. We
don't have a problem with our notepad since we never need to worry
about pops of size larger than $s$, or sequences of more than $s$ pops
belonging to an equivalent pair. We do have a problem with the
alphabet as we seem to need an infinite number of pop
symbols. However, we deal with this by introducing a new 
symbol $\gamma$, whose operational interpretation is as a single pop
of the stack, but which we require to occur in blocks of length
at least $s+1$ (a condition which is easy to enforce while retaining
finitely many states). When $\gamma$-pops are occurring, no recording takes
place because these can never be part of a rise-fall sequence. So,
that leaves only:

\begin{question}
Does the class of permutations generated by an unlimited fork stack
have an algebraic generating function?
\end{question}


\begin{thebibliography}{99}

\bibitem{Atk1} M.~D. Atkinson: Restricted permutations, Discrete
Math. 195 (1999), 27--38.

\bibitem{Atk2} M. D. Atkinson: Generalised stack permutations,
Combinatorics, Probability and Computing 7 (1998), 239-246.

%\bibitem{AMR}
%M.~D. Atkinson, M.~M. Murphy, N. Ru\v{s}kuc: Partially
%well-ordered closed sets of permutations. Technical Report: Otago
%University Department of Computer Science OUCS-2001-05
%(http://www.cs.otago.ac.nz/trseries) (2001) 15 pp.

%\bibitem{Bos}
%P. Bose, J.~F. Buss, A. Lubiw: Pattern matching for
%permutations, Inform. Process. Lett. 65 (1998), 277--283.

\bibitem{ChS} N.~Chomsky and M.~P.~Sch\"{u}tzenberger: The Algebraic
Theory of Context-Free Languages, in:
{\em Computer programming and formal systems} ed. P. Braffort, D. Hirschberg, North Holland, Amsterdam,
1963, 118-161.

\bibitem{FO}
P. Flajolet and A.~M.~Odlyzko: Singularity analysis of generating
functions. SIAM Jour. Disc. Math. 2 (1990), 216-240.

\bibitem{FS}
P. Flajolet and R. Sedgwick: The Average Case Analysis of Algorithms,
Complex Asymptotics and Generating Functions. INRIA Research Report
2026, 1993.

\bibitem{GJ}
I.~P. Goulden, D.~M. Jackson: {\em Combinatorial Enumeration}, John
Wiley and Sons, New York, 1983.

\bibitem{HoU}
J.~E. Hopcroft, J.~D Ullman: {\em Introduction to Automata Theory,
Languages, and Computation}, Addison-Wesley, Reading, Mass. (1979).	

\bibitem{Knu}
D.~E. Knuth: {\em Fundamental Algorithms, The Art of Computer
Programming} Vol. 1 (First Edition), Addison-Wesley, Reading, Mass.
(1967).

%\bibitem{MaV}
%T. Mansour, A. Vainshtein: Restricted 132-avoiding permutations,
%Adv. Appl. Math. 26 (2001) 258-269.


\bibitem{Pra}
V.~R. Pratt: Computing permutations with double-ended
queues, parallel stacks and parallel queues, Proc. ACM Symp. Theory of
Computing 5 (1973), 268--277.


%\bibitem{Sim}
%R. Simion, F.~W. Schmidt: Restricted permutations,
%Europ. J. Combinatorics 6 (1985), 383--406.

\bibitem{Slo}
N.~J.~A. Sloane: {\em The On\-line En\-cyclo\-pedia of In\-teger
Sequences}, \\
\verb+http://www.research.att.com/~njas/sequences/+, 2002.

%\bibitem{Sta1}
%Z.~E. Stankova: Forbidden subsequences, Discrete Math. 132
%(1994), 291--316.

%\bibitem{Sta2}
%Z.~E. Stankova: Classification of forbidden
%subsequences of length $4$, European J. Combin. 17 (1996), no. 5,
%501--517.

\bibitem{Tar}
R.~E. Tarjan: Sorting using networks of queues and stacks, Journal of the
ACM 19 (1972), 341--346.

%\bibitem{Wes}
%J. West: Generating trees and forbidden sequences, Discrete Math. 157
%(1996), 363--374.

\bibitem{Wil}
H.~S. Wilf: {\em generatingfunctionology}, Academic Press, New York, 1993.

\end{thebibliography}
\end{document}